\documentclass[aps,prd,preprint,superscriptaddress,showpacs]{revtex4}
\usepackage{amsmath}
\usepackage{amssymb}
\usepackage{epsfig}
\usepackage{ulem}
\usepackage{graphics}
\usepackage{indentfirst}
\usepackage{color}

\newcounter{RomanNumber}
\newcommand{\MyRoman}[1]{\setcounter{RomanNumber}{#1}\Roman{RomanNumber}}

\begin{document}

\preprint{ACT-4-14, MIFPA-14-15}

\title{Chaotic Inflation in No-Scale Supergravity with String Inspired Moduli Stabilization}

\author{Tianjun Li}
\affiliation{State Key Laboratory of Theoretical Physics
and Kavli Institute for Theoretical Physics China (KITPC),
      Institute of Theoretical Physics, Chinese Academy of Sciences,
Beijing 100190, P. R. China}

\affiliation{School of Physical Electronics,
University of Electronic Science and Technology of China,
Chengdu 610054, P. R. China}

\author{Zhijin Li}

\affiliation{George P. and Cynthia W. Mitchell Institute for
Fundamental Physics and Astronomy,
Texas A\&M University, College Station, TX 77843, USA}

\author{Dimitri V. Nanopoulos}

\affiliation{George P. and Cynthia W. Mitchell Institute for
Fundamental Physics and Astronomy,
Texas A\&M University, College Station, TX 77843, USA}

\affiliation{Astroparticle Physics Group, Houston Advanced
Research Center (HARC), Mitchell Campus, Woodlands, TX 77381, USA}

\affiliation{Academy of Athens, Division of Natural Sciences,
28 Panepistimiou Avenue, Athens 10679, Greece}

\begin{abstract}

The simple chaotic inflation is highly consistent with the BICEP2 experiment, and no-scale supergravity
can be realized naturally in various string compactifications. Thus, we construct a chaotic inflation model
in no-scale supergravity inspired from Type \MyRoman{2}B string compactification with an anomalous $U(1)_X$
gauged symmetry. We introduce two moduli $T_1$ and $T_2$ which transform non-trivially under $U(1)_X$,
and some pairs of fundamental quarks charged under the $SU(N)\times U(1)_X$ gauge group. The non-trivial
transformations of moduli under $U(1)_X$ lead to a moduli-dependent Fayet-Iliopoulos (FI) term.
The modulus $T_2$ and the real component of $T_1$ are stabilized by the non-perturbative effect
from quark condensation
and the $U(1)_X$ D-term. In particular, the stabilization from the anomalous $U(1)_X$ D-term with
moduli-dependent FI term is crucial for inflation
since it gives heavy mass to the real component of the modulus $T_1$ while keeping its axionic part light.
Choosing the proper parameters, we obtain a global Minkowski vacuum where the imaginary part of $T_1$
has a quadratic potential for chaotic inflation.

\end{abstract}


%\pacs{11.25.-w, 98.80.-k}
\maketitle

\section{Introduction}

Inflation is a candidate to solve several problems in the standard big bang model, such as
the horizon problem, flatness problem, large structure of the Universe, etc. And it is getting closer to
be verified based on the recent Planck and BICEP2 observations~\cite{Ade:2013uln, Ade:2014xna}. 
Both experimental results
support the single field inflation with scalar spectral index $n_s$ around $0.96$. However,
the Planck results provide an upper bound on the tensor-to-scalar ratio, $r\leqslant0.11$
at $95\%$ C.L.~\cite{Ade:2013uln}. The simple chaotic inflation model with quadratic potential
$V=\frac{1}{2}m^2\phi^2$, whose $n_s$ is out of this range, is disfavored. In contrast,
the Starobinsky model fits with the Planck data very well~\cite{star}. Consequently,
it was important to realize the Starobinsky model from fundamental
theories, such as the supergravity (SUGRA) theory and string theory before the BICEP2 results.

The no-scale SUGRA~\cite{Cremmer:1983bf}, which can be realized naturally
in various string compactifications~\cite{Witten:1985xb, Li:1997sk}, solves
the cosmological constant problem elegantly.
The Starobinsky model was realized in $SU(2, 1)/U(1)$ no-scale SUGRA with Wess-Zumino superpotential~\cite{ENO}.
Following this development, the SUGRA extensions of the Starobinsky model have been
revived~\cite{Cecotti:1987sa, KT, KL} (For more details and references, see \cite{Linde:2014nna}.).
Besides, the Starobinsky-like inflation can be fulfilled in string theory as well~\cite{Cicoli:2013oba, Ellis:2014cma}.

Very recently, the BICEP2 Collaboration announced the range of tensor-to-scalar ratio based on
the observations of CMB B-mode polarization, $r=0.20^{+0.07}_{-0.05}$ or $r=0.16^{+0.06}_{-0.05}$ by
substracting the dust contributions. Such large $r$ significantly changes the directions of
the inflation model building. The Starobinsky model is now disfavored by the BICEP2 results.
Moreover, many inflation models from string theory predict small $r$ far below $0.01$
and thus contradict with the BICEP2 results~\cite{Burgess:2013sla}. Interestingly,
chaotic inflation is indeed favoured after the BICEP2 results, 
and since has been studied extensively~\cite{Nakayama:2014koa, Harigaya:2014sua, Harigaya:2014qza, 
Ferrara:2014ima, Lee:2014spa, Gong:2014cqa, Ibanez:2014zsa, Ashoorioon:2014nta, Okada:2014lxa,
Kallosh:2014qta, Ellis:2014rxa, Creminelli:2014oaa, Oda:2014rpa, Kaloper:2014zba, 
Hebecker:2014eua, Murayama:2014saa, Farakos:2014gba, Gao:2014fha}.

The no-scale SUGRA is equipped with a curved K\"ahler manifold~\cite{Cremmer:1983bf}, which leads to
non-canonical kinetic terms for the fields along non-flat directions (without shift symmetry).
Normally, the fields with such kind of kinetic terms move too fast toward the minimum,
and then no inflation can be triggered. Alternatively, through parameter tuning it is possible to
get a flat direction for inflation with very small $r$, 
such as the no-scale Starobinsky model~\cite{ENO}. In short,
these potentials are either too steep or too flat to generate chaotic inflation.

For chaotic inflation, the inflaton is preferred to be the scalar with a flat direction on the K\"ahler manifold,
while all the other fields along the rest of the 
directions should be properly stabilized. The flat directions
of the K\"ahler manifold are guaranteed by the shift symmetries of the K\"ahler potential. The scalar potential
is also flat if the shift symmetry is not broken by the superpotential. Thus, the shift symmetry was employed
to construct the chaotic inflation model in no-scale SUGRA~\cite{Ferrara:2014ima}
(For a related study, see~\cite{Ellis:2014rxa}.).
This work is based on the SUGRA extension of the Starobinsky model, the K\"ahler potential is
of no-scale $SU(2, 1)/U(1)$ type, and the inflaton is the imaginary part of the modulus,
which preserves the exact shift symmetry. However, the real component of
the modulus is not stabilized during inflation since the masses of the real component and inflaton
are comparable around the same scale. The point is that the shift symmetry is broken by the superpotenital
explicitly,  in consequence there is no symmetry that can prevent the inflaton from obtaining heavy mass.

Besides the shift symmetry,  moduli stabilization is also needed for chaotic inflation. The moduli can be
stabilized by the non-perturbative effects~\cite{Kachru:2003aw} via the KKLT mechanism
in an anti-de Sitter (AdS) vacuum, which is uplifted to a 
metastable de Sitter (dS) vacuum by sets of anti D3-branes
where supersymmetry (SUSY) is broken explicitly. Burgess, Kallosh, and Quevedo (BKQ) suggested that
the uplifting of the AdS vacua with spontaneously SUSY breaking can be realized by the D-term associated
with an anomalous $U(1)_X$ gauge symmetry~\cite{Burgess:2003ic}. Nonetheless, the non-perturbative part of
the superpotentials in both KKLT and BKQ are not invariant under anomalous $U(1)_X$. The gauge invariant moduli
stabilization was proposed in Refs.~\cite{Dudas:2005vv, Achucarro:2006zf} based on the non-perturbative effect
of hidden gauge symmetry. As the D-term is positive semi-definite, it is useful to construct the Minkowski
or dS vacuum. The effects of the D-term on moduli stabilization and dS vacua are also studied in
Ref.~\cite{Binetruy:2004hh}. Moreover, the F-term is widely used to uplift the AdS vacua.
In Ref.~\cite{Kallosh:2006dv}, the O'Raifeartaigh model with quantum corrections is introduced to
the KKLT scenario. The heavy fields are integrated out while a light field is fixed
at very small value. Its F-term contributes to the vacuum energy and uplift the AdS vacua obtained
from KKLT. Similarly, the Polonyi model can uplift the AdS vacua when combined with the KKLT mechanism~\cite{Abe:2006xp}.

The anomalous D-term with moduli-dependent Fayet-Iliopoulos (FI) term plays a special role in inflation. 
The FI term depends on the real component of the moduli only, so stabilization through such kind of
D-term only gives heavy mass to the real component while the imaginary or 
axion-like part remains light. This is different
from the stabilization by the F-term or D-term with constant FI term, in which cases both the real
and imaginary components appear in the potentials and it is difficult to separate the masses
between the real and imaginary parts at different scales.
Instead of stabilizing the moduli directly,  chaotic-like inflation can also be obtained in no-scale SUGRA
by minimizing a term combining the moduli and matter fields~\cite{pr-no-scale}, nevertheless, the moduli
are indeed not stabilized during inflation.

In this work, we will apply the shift symmetry of the moduli to obtain
 chaotic inflation, where the inflaton is
the corresponding axion-like field. In particular, the shift symmetry is preserved by both in K\"ahler potential
and superpotential. So it can be consistently gauged to form the anomalous $U(1)_X$ as long as the gauge anomaly
is cancelled. The K\"ahler potential is inspired from the Type \MyRoman{2}B string compactification, where
two moduli $T_1$ and $T_2$ are charged under the anomalous $U(1)_X$ gauge symmetry.
One of the moduli $T_2$ is stabilized by the KKLT mechanism in a gauge invariant way.
The real component of $T_1$ is automatically stabilized by the D-term associated with
the anomalous $U(1)_X$, while its imaginary component remains light and is a natural candidate
for the inflaton. We choose the moduli stabilization scale at least one order of magnitude
 higher than
the inflation scale so that the inflation and moduli stabilization can be separated into two stages.

This paper is organized as follows. In Section 2, we briefly review gauge invariant moduli stabilization
based on non-perturbative effects. In Section 3, we discuss the anomaly cancellation of the anomalous $U(1)_X$
symmetry. In Section 4, we show by combining the non-perturbative effects and the D-term, all the moduli
except the axion-like imaginary component of $T_1$ are stabilized. Choosing the proper parameters, we get
the Minkowski vacuum and a light axion-like field with quadratic potential, which generates chaotic inflation
in the scale far below the moduli stabilization scale. We discuss model building and 
then conclude in Section 5.

\section{Gauge Invariant Moduli Stabilization}

In the KKLT proposal, the dilaton and complex-structure moduli of Calabi-Yau compactification
are fixed by the backgound NSNS and RR fluxes. Thus, there is only one K\"ahler modulus, $T$, which
 is not fixed by the fluxes. The SUGRA description of its low-energy effective theory is given
by the K\"ahler potential
\begin{equation}
K=-3\log (T+\bar{T})~,
\end{equation}
and the superpotential
 \begin{equation}
 W=W_0+W_{np}~,
 \end{equation}
where the constant term $W_0$ is obtained from the fluxes which are used to stabilize
the dilaton and complex-structure moduli, and the non-perturbative term $W_{np}=Ae^{-aT}$ is
generated by the Euclidean D3-branes or alternatively by gaugino condensation within
a non-Abelian sector from a stack of wrapped D7-branes.
The generic F-term scalar potential is given by
\begin{equation}
V_F=e^K(K^{i\bar{j}}D_iW D_{\bar{j}}\bar{W}-3W\bar{W})~, \label{vf}
\end{equation}
in which $K^{i\bar{j}}$ is the inverse of the K\"ahler metric $K_{i\bar{j}}=\partial_i\partial_{\bar{j}}K$. The potential of the modulus $T$ admits a supersymmetric AdS vacua where $T$ is stabilized. There are several ways to uplift the AdS vacua to dS vacua. In the original KKLT proposal, the AdS vacuum is uplifted by anti D3-branes,
which generates a non-supersymmetric term in the scalar potential
\begin{equation}
V=V_F+\frac{D}{\sigma^2}~,
\end{equation}
where $\sigma=\rm{Re}(T)$, and $D$ is a constant.

Instead of breaking SUSY explicitly, the AdS vacua can be uplifted by the D-term in the
BKQ proposal~\cite{Burgess:2003ic}. In general, the D-term for the four-dimensional
 $N=1$ gauged SUGRA with K\"ahler term $G=K(\phi,\bar{\phi})+\log (W\bar{W})$ is
\begin{equation}
V_D=\frac{1}{2}D_aD^a~,
\end{equation}
where the gauge indices are raised by the form $[(Ref)^{-1}]^{ab}$ with $f$
 the gauge kinetic function. The $D_a$ components are
\begin{equation}
D_a=iK_iX^i_a+i\frac{W_i}{W}X^i_a~,
\end{equation}
or
\begin{equation}
D_a=iK_iX^i_a~,
\end{equation}
if $W$ is gauge invariant. Here, the $X^i_a$ are the
components of the Killing vector $X_a=X^i_a(\phi)\partial/\partial\phi^i$,

In the BKQ proposal, the Killing vector has components $X^T=\frac{2E}{3}i$ and $X^{Q_i}=iq_iQ_i$.
So the D-term is
\begin{equation}
V_D=\frac{g^2_{YM}}{2}D^2=\frac{2\pi}{\sigma}(\frac{E}{\sigma}+\sum q_i |Q_i|^2)^2,
\end{equation}
where $g^2_{YM}=4\pi/\sigma$, $Q_i$ are the matter fields which
transform linearly under the anomalous $U(1)_X$ with charges $q_i$, and
the modulus $T$ shifts under the anomalous $U(1)_X$ and
then is related to a field dependent FI term. 
That the D-term should be non-vanishing (non-cancellability) is a critical assumption to
uplift the AdS vacua.
It was argued that the matter fields $Q_i$ can obtain vacuum expectation value (VEV) $\langle Q_i \rangle=0$,
so the D-term is similar to the effect of the anti D3-branes. The uplift of AdS vacua can
also be done by F-term (For example, see~\cite{Lebedev:2006qq, Dudas:2006gr}.).

As noticed in the BKQ proposal and Ref.~\cite{Choi:2005ge}, it is not consistent to 
directly add the D-term in KKLT mechanism. The modulus $T$ shifts under anomalous $U(1)_X$: $T\rightarrow T+X^T\epsilon$, where $X^T$ is the Killing vector generating $U(1)_X$ transformation of modulus $T$. The non-perturbative term in the superpotential $W_{np}$ is not gauge invariant: $W_{np}\rightarrow e^{-aX^T\epsilon}W_{np}$. So 
 a field dependent coefficient in the non-perturbative term is required
 to cancel the factor $e^{-aX^T\epsilon}$ if the shift symmetry can be gauged consistently.

A gauge invariant non-perturbative superpotential was constructed in 
Refs.~\cite{Dudas:2005vv, Achucarro:2006zf}. The $W_{np}=Ae^{-aT}$ in KKLT is 
replaced by $W_{np}=F(Q_i)e^{-aT}$, in which $F(Q_i)$ is product of matter fields $Q_i$.
 The $F(Q_i)$ transformation under $U(1)_X$ cancels the phase factor $e^{-aX^T\epsilon}$ so that
 the new $W_{np}$ is indeed invariant under $U(1)_X$. Such kind of
 superpotential, which originated from gaugino condensation,
 has been studied in Refs.~\cite{Taylor:1982bp, Affleck:1983mk, Lust:1990zi, deCarlos:1991gq}.
Specifically, the model employs $N_0$ fundamental quark pairs, $(Q_i, \bar{Q_i})$ under
 gauge group $SU(N)\times U(1)_X$, and the $U(1)_X$ charges of the quark pairs are $(q, \bar{q})$.
The quarks condense and form the composite meson fields $|M|^2=|M_i|^2$, in which $M^2_i=Q^i\bar{Q}_i$ for $i=1, \cdots, N_0$. Consequently, the effective superpotential after the condensation is
\begin{equation}
W_{np}=(N-N_0)M^{-\frac{2N_0}{(N-N_0)}}e^{\frac{(q+\bar{q})N_0T}{(N-N_0)\delta_{GS}}},
\end{equation}
in which the $i\delta_{GS}=X^T$ and its value is determined by the quantum anomaly cancellation conditions for $SU(N)^2\times U(1)_X$ and $U(1)^3_X$, which will be discussed later. Obviously, the above superpotential is gauge invariant.
Opposite signs are assigned for $q+\bar{q}$ and $\delta_{GS}$. The $U(1)_X$ D-term is
\begin{equation}
V_D\propto (N_0(q+\bar{q})|M|^2-\frac{3\delta_{GS}}{2\sigma})^2.
\end{equation}
This D-term is non-vanishing for $\sigma<\infty$ and its minimum is located
 at $\langle M\rangle=0$. The non-cancellability assumption, which is crucial for the BKQ proposal, now is realized in the condensation mechanism. However, the non-cancellability directly results
 from the fact that $\delta_{GS}/(q+\bar{q})<0$, by introducing more fields charged under $U(1)_X$, this non-cancellability disappears.

The modulus $T$ and the composite field $M$ can be stabilized by minimizing $V_F$ or the combination $V_F+V_D$. 
Based on purely $V_F$, the vacua are of AdS as usual, uplifting from the D-term results in the dS vacua.

\section{Anomaly Cancellation of Anomalous $U(1)_X$}

Anomalous $U(1)_X$ symmetry is obtained in the
heterotic string by gauging the shift symmetry of the axion-dilaton multiplet
$S$~\cite{Dine:1987xk}. The gauge kinetic term is
\begin{equation}
\int d^2\theta fW^2_{\alpha}~,
\end{equation}
where $W_\alpha$ is the field strength of the $U(1)_X$ vector superfield,
and the gauge kinetic function is taken as $f=S$. The gauge kinetic term contains two parts $Re(f)F^2$ and $Im(f)F\tilde{F}$ \cite{Wess:1992cp}.  The second term has a non-trivial transformation of $S$ and plays a
crucial role in gauge anomaly cancellation through the Green-Schwarz mechanism in four-dimensional
 spacetime \cite{Green:1984sg}. The quantum anomaly of $U(1)_X$ is cancelled by the term introduced from the transformation $S\rightarrow S+i\epsilon\delta_{GS}$. However, for the heterotic string case, the anomalous $U(1)_X$ 
is very constrained. As argued in Ref.~\cite{Dine:1987xk}, only the superfield $S$ can be transformed non-trivially under anomalous $U(1)_X$, otherwise there will be unwanted mass terms and tadpoles at tree level. The FI terms introduced by $S$ appear in heterotic string at higher loop levels, so they are expected to be much smaller than the tree-level potential.
In consequence, they are not useful if a large D-term is needed.

In Type \MyRoman{2}B string compactification, generally there are several moduli, $T_i$, from the Calabi-Yau space. The moduli-dependent part of the gauge kinetic function is $f=g^i_aT_i$, where $g^i_a$ are positive constants.
Anomaly cancellation is determined by the component $g^i_a Im(T_i)F_a\tilde{F}_a$. Given the moduli $T_i$ transform as $T_i\rightarrow T_i+i\delta^i_a \epsilon$ under $U(1)_a$, the anomaly cancellation requires
\begin{equation}
\sum g^i_a\delta^i_a=-\Delta,
\end{equation}
where  $\Delta$ is the coefficient of the gauge anomaly from the fermionic contributions $\Delta F_a\tilde{F}_a$. Specifically, for the $U(1)^3_X$ gauge anomaly, the above anomaly cancellation turns into
\begin{equation}
\sum g^i_X\delta^i_X=-\frac{1}{48\pi^2}\sum q^3_m, \label{ano}
\end{equation}
in which $q_m$ are the charges of quarks, and a factor $1/3$ is 
attributed to the over-counting of the anomaly diagrams.

\section{ Chaotic Inflation Model Building}

From the above discussions, the anomalous $U(1)_X$ in Type \MyRoman{2}B string theory
instead of heterotic string theory is preferred. In particular, in the Type \MyRoman{2}A intersecting
D6-brane model building or its T-dual Type \MyRoman{2}B D3-D7 brane model building,
we will not only have up to four anomalous $U(1)$ gauge symmetries, but also have the hidden sector
with additional gauge groups and exotic particles~\cite{Cvetic:2001tj, Cvetic:2001nr, 
Cvetic:2004ui, Chen:2006gd, Chen:2007af}. Inspired by these string constructions,
we consider the following K\"ahler potential
\begin{equation}
K=-\log (T_1+\bar{T_1})-2\log(T_2+\bar{T_2})+\sum_{i=1}^{N_0}(Q_i\bar{Q_i}+\tilde{Q_i}\bar{\tilde{Q_i}})+S\bar{S}-\frac{(S\bar{S})^2}{\Lambda^2_1}+X\bar{X}-\frac{(X \bar{X})}{\Lambda^2_2}, \label{kh}
\end{equation}
where $T_1$ and $T_2$ transform non-trivially, $T_i\rightarrow T_i+i\delta^i_{GS}\epsilon$ under the anomalous $U(1)_X$, the $N_0$ quark pairs $(Q_i, \tilde{Q_i})$ form
 fundamental representation of $SU(N)$ gauge symmetry with $U(1)_X$ charges $(q, \tilde{q})$. As proposed before, the quarks $Q_i$ condense and form composite meson fields $M_i=\sqrt{Q^i\tilde{Q_i}}$. The superfields $S$ and $X$ are neutral under $SU(N)\times U(1)_X$, and the higher order terms $(S\bar{S})^2/\Lambda^2_1$ and $(X \bar{X})^2/\Lambda^2_2$ from quantum corrections are needed to fix $S$ and $X$ at $\langle S\rangle=\langle X\rangle=0$ during inflation.
 $S$ is from the O'Raifeartaigh model and used to uplift the AdS vacua in the KKLT mechanism~\cite{Kallosh:2006dv},
while $X$ provides the non-vanishing F-term for inflation.

The kinetic terms of the fields $\phi^i\equiv(T_i, Q_i, \tilde{Q_i}, S, X)$ are given by $L_{kin}=K_{i\bar{j}}\partial_\mu \phi^i \partial^\mu \bar{\phi}^{\bar{j}}$ with the K\"ahler metric $K_{i\bar{j}}\equiv\partial^2K/\partial\phi^i\partial\bar{\phi}^{\bar{j}}$. From the K\"ahler potential in Eq.~(\ref{kh}), fields $Q_i, \tilde{Q_i}$, $S$,
 and $X$ (at lowest level) have canonical kinetic terms, while for the moduli $T_i$ with no-scale type K\"ahler potential, their kinetic terms are
\begin{equation}
L_K=\frac{\partial_\mu T_1 \partial^\mu \bar{T_1}}{(T_1+\bar{T_1})^2}+\frac{2\partial_\mu T_2 \partial^\mu \bar{T_2}}{(T_2+\bar{T_2})^2}~. \label{mkn}
\end{equation}

The gauge kinetic term consists of two parts, $SU(N)$ and $U(1)_X$. Here, we focus on the $U(1)_X$
due to quark condensations. The $SU(N)^2\times U(1)_X$ gauge anomalies are cancelled by the shifts of
the gauge kinetic function $f_{SU(N)}\propto g_aT_2$ under the anomalous $U(1)_X$.
The gauge kinetic term of $U(1)_X$ is
\begin{equation}
\int d^2\theta (g_1T_1+g_2T_2)W^2_{\alpha}~.
\end{equation}

The parameters $\delta^i_{GS}, ~g_1$, and $g_2$ are free as along as the anomaly cancellation conditions
 are satisfied. Here we take
\begin{equation}
\delta\equiv\delta^2_{GS}=-\delta^1_{GS}, \label{des}
\end{equation}
and $g_1=1, g_2=2$. The anomaly cancellation condition in Eq.~(\ref{ano}) gives
\begin{equation}
\delta=-\frac{NN_0(q^3+\bar{q}^3)}{48\pi^2}~.~\,
\end{equation}

The superpotential of the gauged SUGRA, first of all, should be gauge invariant. If there is only one modulus transforms non-trivially under anomalous $U(1)_X$, then the formula of the superpotential is strongly constraint by the gauge invariance. There is only one choice $W(T)\sim e^{aT}$, just the effective superpotential from non-perturbative effects. However, by employing two moduli ($T_1$, $T_2$) with Killing vector
\begin{equation}
 X^{T_1}=i\delta_a, ~~X^{T_2}=-i\delta_b,
\end{equation}
the constraint is relaxed. The combination  of the two moduli $\delta_bT_1+\delta_aT_2$ is automatically gauge invariant, and then any function in terms of $\delta_bT_1+\delta_aT_2$ is gauge invariant. This is crucial to construct the gauge invariant superpotential.

We consider the following superpotential
\begin{equation}
W=w_0+(N-N_0)M^{-\frac{2N_0}{N-N_0}}e^{\frac{N_0(q+\tilde{q})T_2}{\delta (N-N_0)}}-\mu S+aX(T_1+T_2+s)~,
\end{equation}
where the first two terms ($W_{st}$) are for the gauge invariant stabilization of modulus $T_2$. The third term is to uplift the AdS vacuum, while the last term ($W_{in}$) is to generate chaotic inflation, and $s$ is a constant.
For the moduli stabilization, we require that $W_{st}$  be ``hierarchically" larger than the term $W_{in}$.
To be concrete, we will take  $w_0\simeq 2.0 \times 10^{-3}$
in Planck units, while the parameter $a$, which corresponds to the inflaton mass, is about $10^{13}$ GeV,
or $\sim 10^{-5}$ in Planck units. Therefore, the term $W_{in}$ has ignorable effect on the moduli stabilization.
Conversely, once the moduli are fixed, they are completely frozen out during inflation.

\begin{figure}
\centering
\includegraphics[width=110mm, height=55mm,angle=0]{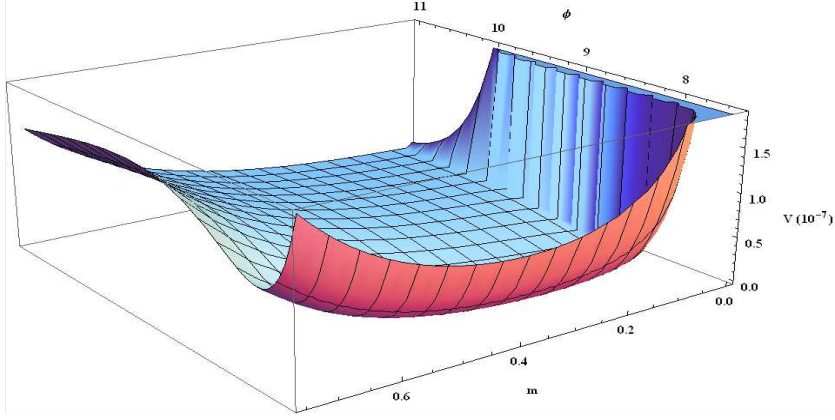}
\caption{ {\small F-term moduli stabilization in a Minkowski vacuum.}}
\label{fig1}
\end{figure}

Given $T_2\equiv\phi+i\theta$, $M\equiv m e^{i\beta}$ and $a=-\frac{N_0(q+\tilde{q})}{\delta (N-N_0)}$, $b=\frac{2N_0}{N-N_0}$, the F-term potential is
\begin{equation}
\begin{split}
V_F=\frac{e^{2N_0m^2}}{4(T_1+\bar{T_1})\phi^2}\{e^{-2a\phi}[4(N-N_0)^2m^{-2b}(\frac{1}{2}a^2\phi^2+a\phi) ~~~~~~~~~~~~~~~~~~~~~~~~~~~~~~\\
+\frac{(N-N_0)^2}{2N_0}b^2m^{-2(b+1)}-2b(N-N_0)^2m^{-2b}+2N_0m^{2-2b}(N-N_0)^2)]~\\
+2N_0m^2w^2_0+2w_0(N-N_0)m^{-b}e^{-a\phi}(2a\phi+2N_0m^2-b){\rm{cos}}(a\theta+b\beta) \\
+\mu^2+a^2|T_1+T_2+s|^2+\cdots\},\label{fp}~~~~~~~~~~~~~~~~~~~~~~~~~~~~~~~~~~~~~~~~~~~~~~~~~
\end{split}
\end{equation}
in which the terms proportional to $S$ and $X$ are ignored. The potential depends on
the combination $a\theta+b\beta$, and has a flat direction, the Goldstone boson which becomes the longitudinal component of the $U(1)_X$ massive vector field through the Higgs mechanism.

From the potential in Eq.~(\ref{fp}), the term $\propto a^2$ is several orders smaller than
 the KKLT terms and only has small correction to the moduli stabilization. Its effect will be studied later.
The modulus $T_2$ and meson $M$ are stabilized by the KKLT term, which gives an AdS vacuum.
The AdS vacuum is raised by the term $\propto\mu^2$. In our model, taking $N=10$ with only one flavor of quarks,
 $q+\tilde{q}=-4\pi\delta$, the constant $w_0=0.002$, the vacuum locates
at $\phi_0=8.2990$, $m_0=0.0879$ and $a\theta+b\beta=(2n+1)\pi$. To uplift the AdS vacuum to a Minkowski vacuum,
we require $\mu^2=6.58\times 10^{-6}$.
The masses of the fields $\phi$ and $m$ (after canonical normalization) 
are $m_\phi=1.2\times 10^{-3}$ and $m_m=1.3\times 10^{-4}$ in Planck units.
For $r\simeq 0.16$, we have the Hubble scale around $5\times 10^{-5}$ in Planck units.
Thus, both $\phi$ and $m$ can be stabilized during inflation. Also,
the gravitino mass is $4.5\times 10^{-5}$, which will not affect the inflation in no-scale supergravity.
Moreover, we present the potential for moduli stabilization in Fig.~\ref{fig1}.

The D-term associated with the anomalous $U(1)_X$ is
\begin{equation}
V_D=\frac{1}{2Re(T_1+2T_2)}(\frac{\delta}{T_1+\bar{T_1}}-\frac{2\delta}{T_2+\bar{T_2}}+N_f(q+\tilde{q})|M|^2)^2~. \label{vd}
\end{equation}
The D-term for another gauge group $SU(N)$ has already vanished under the quark condensation $|Q_i|^2=|\tilde{Q}_i|^2$.
In Eq.~(\ref{vd}) the $\delta$ has opposite sign to the charge $q+\tilde{q}$, the above D-term is cancelled
 by shifting the real component of modulus $T_1$ for any given $T_2$ and $M$.
This is completely different from the case with modulus $T_2$ only, in which the $|M|^2$ has the same sign
with the modulus-dependent FI term and non-cancellability is guaranteed. At the vacuum,
 the D-term vanishes, and gives a large mass to the field ${\rm Re}(T_1)$. So even though we can
gauge invariantly fix the modulus $T_2$ and $M$, the D-term uplifting of the AdS vacua is not feasible.

The modulus $T_2$ and $M$ are fixed at $\langle T_2 \rangle=\phi_0+i\theta_0$ and $|M|=m_0$.
For simplicity, we take the $U(1)_X$ gauge $\theta_0=0$. The real component of modulus $T_1$ obtains
a large mass and is stabilized as well. For $T_1=\sigma+i\rho$, the vacuum locates at
\begin{equation}
\sigma_0=\frac{1}{2}\left(\frac{1}{\phi_0}+|\frac{q+\tilde{q}}{\delta}|N_0m^2_0 \right)^{-1},
\end{equation}
with mass (before rescaling)
\begin{equation}
m^2_\sigma=\left.\frac{\partial^2 V_D}{\partial\sigma^2}\right|_{\sigma_0, \langle T_2\rangle, m_0}=\frac{8\delta^2}{\phi^5_0}\frac{(1-N_0(q+\tilde{q})\phi_0m^2_0/\delta)^5}{5-4N_0(q+\tilde{q})\phi_0m^2_0/\delta}~.
\end{equation}
So the mass of $\sigma$ seems to be strongly depending on the modulus through $\phi_0^{-5}$.
However, it can be easily compensated by modifying the negative ratio $(q+\tilde{q})/\delta$,
and gives a large mass of modulus $\sigma$ with $m_\sigma\sim O(M_p)$.

Here the modulus-dependent FI term plays a crucial role in the moduli stabilization, as it is independent
with the imaginary components of the moduli, we can safely stabilize the real component while keep
the axion-like imaginary component light.

\begin{figure}
\centering
\includegraphics[width=90mm, height=60mm,angle=0]{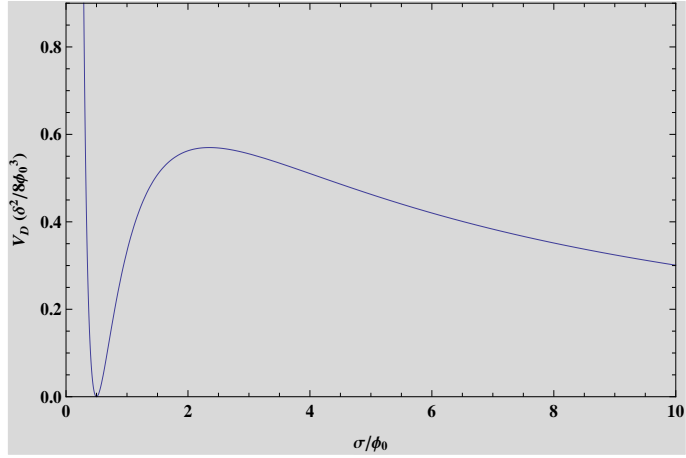}
\caption{ {\small Anomalous $U(1)_X$ D-term with unit $\frac{\delta^2}{8\phi^3_0}$. The $x$ axis
is scaled by $\phi_0$, and the quark condensation term is ignored.
With non-vanished $m^2_0$ term the $\langle \sigma \rangle$ will shift to the left and
makes the minimum valley steeper. }}
\label{fig2}
\end{figure}

Fig. \ref{fig2} shows the D-term potential with stabilized $T_2$, where the quark condensation term is ignored.
The field $\sigma$ has a steep minimum at $\phi_0/2$, which is also the global minimum
as a result of the cancellability. Besides, the potential shows run away tendency corresponding to
 the decompactification. Considering the meson contribution,
the minimum will shift to the left, in such case the minimum valley gets steeper
and gives a stronger stabilization.

The higher order term in F-term potential gives a small correction to the $T_1$ stabilization.
The overall potential near the vacuum is
\begin{equation}
\begin{split}
V=V_F|_{\phi_0, m_0}+V_D\simeq\frac{a^2}{8\phi^2_0\sigma_0}\rho^2+f(\sigma)+\frac{1}{2}m^2_\sigma(\sigma-\sigma_0)^2  ~~~~~~~~~~~~~~~~~~~~~~~~\,~~~~\\
\simeq\frac{a^2}{8\phi^2_0\sigma_0}\rho^2+f(\sigma_0)+f'(\sigma_0)(\sigma-\sigma_0)+\frac{1}{2}m^2_\sigma(\sigma-\sigma_0)^2 ~~~~\,\\
=\frac{a^2}{8\phi^2_0\sigma_0}\rho^2+\frac{1}{2}m^2_\sigma(\sigma-\sigma_0+\frac{f'(\sigma_0)}{m^2_\sigma})^2+f(\sigma_0)
-\frac{f'(\sigma_0)^2}{2m^2_\sigma}~, \label{vin}
\end{split}
\end{equation}
in which $f(\sigma)=\frac{a^2}{8\phi^2_0\sigma}(\sigma +\phi_0+s)^2$ and the F-term coefficient $e^{2N_0m^2_0}\simeq1$ is ignored. The Minkowski vacuum is realized for $s=-(\sigma_0+\phi_0)$. Besides, the VEV
 of $\sigma$ is shifted by a tiny part $f'(\sigma_0)/m^2_\sigma\propto a^2\sim 10^{-10}$.

From Eq.~(\ref{vin}), we get the scalar potential for the only non-fixed scalar $\rho$
\begin{equation}
V_{in}(\rho)=\frac{a^2}{8\sigma_0\phi^2_0}\rho^2~.
\end{equation}
And its kinetic term in Eq.~(\ref{mkn}) is
\begin{equation}
L_K=\frac{1}{4\sigma^2_0}\partial_\mu \rho \partial^\mu \rho~.
\end{equation}
Redefining the field $\psi\equiv\rho/\sqrt{2}\sigma_0$, the Lagrangian for the canonically normalized field $\psi$ is
\begin{equation}
L= \frac{1}{2}\partial_\mu \psi \partial^\mu \psi+\frac{\sigma_0}{4\phi^2_0}a^2\psi^2~.
\end{equation}

Chaotic inflation can be driven by a scalar $\psi$ with a quadratic potential, 
which is known to be consistent with the BICEP2 observations, especially for the large tensor-to-scalar ratio $r\simeq \frac{8}{N_e}$, where $N_e$ is the e-folding number of the Universe scale expansion during inflation process. To be consistent with the observations,
the inflaton mass is about $m_\psi\simeq 1.8\times 10^{13}$~GeV. Therefore, the parameter
$a=(2/\sigma_0)^{1/2}\phi_0 m_\psi\sim 10^{-5}$ in Planck unit, as discussed before.

\section{Discussions and Conclusion}

In this work we have constructed the chaotic inflation model in the no-scale SUGRA inspired
from Type \MyRoman{2}B string compactification. The inflation models in no-scale SUGRA generically
give a small tensor-to-scalar ratio $r<0.01$~\cite{ENO, KL, Linde:2014nna}, which are strongly
disfavored by the recent BICEP2 observations~\cite{Ade:2014xna}. For a lot of stringy inflation models,
they are realized from the string low-energy effective actions, which are of no-scale type and
obtain small $r$ as well~\cite{Burgess:2013sla}. The inflations with small $r$ are driven by the scalars
which are non-flat directions on the K\"ahler manifold. The potentials of the these scalar fields are
either too steep for inflation, or of plateau type with small $r$ after tuning. Therefore,
 as correctly noticed in \cite{Ferrara:2014ima}, it is necessary to employ the fields which
are flat directions of K\"ahler manifold. The K\"ahler potential is invariant under the shift of such fields.

However, having only shift symmetry does not guarantee inflation. The extra moduli except the inflaton
should be frozen during inflation to generate single field inflation. The moduli can be stabilized
by non-perturbative effects like the KKLT mechanism. However, once the extra moduli are stabilized,
the inflaton, which has shift symmetry, also obtains mass at the same scale and then destroys
the inflation~\cite{Ferrara:2014ima, Kallosh:2007ig}. In short, the inflaton with shift symmetry
in the K\"ahler potential does not have light mass as expected. The point is that
 the shift symmetry provided in the K\"ahler potential $K$ is broken by the superpotential $W$ explicitly.
To obtain a light modulus, the shift symmetry should be kept 
in the whole K\"ahler function $G=K+\log (W\bar{W})$.

If there is just one modulus $T$ and an anomalous $U(1)_X$ gauge symmetry,
the only modulus-dependent superpotential, which
is invariant up to a phase factor under shift transformation, is $e^{-aT}$. So this is just
the effective superpotential from
 the non-perturbative effects. However, it is impossible to get a quadratic potential
for the chaotic inflation with such a superpotential. In this work, we have
 solved this problem by using two moduli that transform non-trivially under the $U(1)_X$, so that
we can construct a polynomial gauge invariant superpotential.

The KKLT proposal also needs to be modified for the anomalous $U(1)_X$, as in the inital case
the non-perturbative superpotential is not invariant under the anomalous $U(1)_X$~\cite{Kachru:2003aw}.
This is solved by introducing a hidden gauge sector $SU(N)$
gauge group~\cite{Dudas:2005vv, Achucarro:2006zf}. The non-perturbative superpotential obtained from the quark condensation is invariant under $SU(N)\times U(1)_X$, and leads to the moduli stabilization. It also solves the non-cancellability assumption in the BKQ proposal~\cite{Burgess:2003ic}. The moduli stabilization in our work follows this gauge invariant method, but with different role the D-term plays.

In our model we have considered two moduli $T_i$ transforming non-trivially under anomalous $U(1)_X$. It could be obtained
from Type \MyRoman{2}B string compactification instead of the heterotic string compactification since in the later case only
the dilaton superfield can be gauged under anomalous $U(1)_X$ \cite{Dine:1987xk}. Besides,
the Type \MyRoman{2}B string compactification is also preferred as it allows FI term at tree level, the large D-term is
needed to stabilize the moduli at string scale. We have stabilized one of the moduli $T_2$ by the gauge invariant
non-perturbative superpotential. However, differently from Ref.~\cite{Achucarro:2006zf}, 
our D-term vanishes at the vacuum. 

The cancellation of the D-term fixes the real component of another modulus $T_1$, whose
 mass can be at least one-order of magnitude larger than the Hubble scale by changing the $U(1)_X$ charges.
Nonetheless, for the axion-like component of $T_1$, its mass is not affected
by the D-term, and keeps light after the moduli stabilization.
While for the stabilization determined by the F-term or D-term with constant FI term, it is very difficult to
get a light mass after stabilization as all the components interact with each other.

The quadratic potential of the axion-like component is from the F-term of field $X$
from the simplest $U(1)_X$ invariant superpotential term $X(T_1+T_2+s)$. However,
it is easy to get the polynomial potential by adopting the superpotential $Xf(T_1+T_2)$,
with $f$ a general polynomial function.

\begin{acknowledgments}

Z.L would like to thank Ergin Sezgin for valuable discussion. The work of DVN was supported
in part by the DOE grant DE-FG03-95-ER-40917. The work of TL is supported in part by
by the Natural Science
Foundation of China under grant numbers 10821504, 11075194, 11135003, and 11275246, and by the National
Basic Research Program of China (973 Program) under grant number 2010CB833000.

\end{acknowledgments}

\end{document}